\documentclass[showpacs,preprintnumbers,amsmath,amssymb,preprint,aps]{revtex4}
\usepackage[dvips]{graphicx}
\usepackage{graphicx}
\usepackage{amsfonts}
\usepackage{bm}
\usepackage{amsmath}
\usepackage{amssymb}
\usepackage{color}
\usepackage[all]{xy}

\def\be{\begin{equation}}
\def\ee{\end{equation}}
\def\bea{\begin{eqnarray}}
\def\eea{\end{eqnarray}}

\begin{document}

\title{Geometrothermodynamics of phantom  AdS black holes}

\author{Hernando  Quevedo$^{1,2}$, Mar\'\i a N. Quevedo$^3$ and Alberto S\'anchez$^4$}
\email{quevedo@nucleares.unam.mx,maria.quevedo@unimilitar.edu.co,asanchez@nucleares.unam.mx}
\affiliation{
$^1$Instituto de Ciencias Nucleares, Universidad Nacional Aut\'onoma de M\'exico,
 AP 70543, M\'exico, DF 04510, Mexico\\
$^2$Dipartimento di Fisica and ICRANet, Universit\`a di Roma ``La Sapienza",  I-00185 Roma, Italy\\
$^3$ Departamento de Matem\'aticas, 
 Facultad de Ciencias B\'asicas, 
Universidad Militar Nueva Granada, Cra 11 No. 101-80, 
Bogot\'a D.E., Colombia \\
$^4$Departamento de Posgrado, CIIDET,  AP752, 
Quer\'etaro, QRO 76000, Mexico}

\date{\today}

\begin{abstract}

We show that to investigate the thermodynamic properties of charged phantom spherically symmetric anti-de-Sitter black holes, 
it is necessary to consider the cosmological constant as a thermodynamic variable so that the corresponding fundamental equation is a homogeneous function defined 
on an extended equilibrium space. 
We explore all the thermodynamic properties of this class of black holes by using the classical physical approach, based upon the analysis of the fundamental equation, 
and the alternative mathematical approach as proposed in geometrothermodynamics. We show that both approaches are compatible and lead to equivalent results.


\end{abstract}


\maketitle

\section{Introduction}
\label{sec:int}

In theoretical physics, phantom fields are obtained usually as solutions to field equations which follow from the variation of Lagrangian densities with negative kinetic terms. This kind of exotic fields have been investigated for many years, especially in the context of electromagnetic and scalar fields \cite{gibrad96}.
Although the study of phantom fields could seem to have only a pure theoretical motivation, recent cosmological observations have increased the interest in investigating their physical properties. Indeed, since the discovery of the acceleration of the Universe many different models have been proposed to explain such 
an unexpected behavior. One the main characteristics of the Universe acceleration is that it can be explained by assuming the existence of an effective field 
generating repulsive gravity between the elements of the Universe. A repulsive field would correspond in general relativity to a fluid with negative pressure, for instance. This is the main characteristic of the cosmological constant which is an essential ingredient of the $\Lambda$CDM model, the most popular cosmological model in Einstein gravity
\cite{lcdm}. But this is not the only available explanation in Einstein gravity. If repulsive gravity is assumed to be also a part of general relativity, an alternative model can be developed in which no exotic matter is necessary, but the repulsion appears as a geometric effect that affects the evolution \cite{lq15}. 
An alternative source of repulsive gravity in the context of exotic fields would be represented by a distribution with negative energy density, i.e., a phantom field. In fact, comparison with observational data \cite{hann06,dun09} suggests that a phantom field can explain the acceleration of our Universe. It is, therefore, very important to understand the physical properties of phantom fields in the framework of gravity theories.

Recently, a particular new phantom solution was obtained in the framework of the Einstein-Maxwell theory with cosmological constant \cite{jrh12}. The solution represents a spherically symmetric black hole with mass, cosmological constant and charge, as independent physical parameters. The first two parameters come from classical general relativity, but the charge is phantom. This is an important point because the phantom nature of the charge changes drastically the geometric and 
physical properties of the corresponding spacetime, as clearly shown in \cite{jrh12}. In particular, the thermodynamic properties of the new black hole were investigated by using two different approaches, namely, the classical physical approach of black hole thermodynamics \cite{dav77} and the mathematical approach of geometrothermodynamics (GTD) \cite{quev07}. In the special case of this phantom black hole, it was shown in \cite{jrh12}
 that these two approaches lead to different results. 
In particular, the points where phase transitions occur, as predicted by GTD, do not coincide with the phase transition structure found by analyzing the behavior of the corresponding heat capacity, which is the essential variable used in classical black hole thermodynamics in order to determine phase transitions. Then, the authors 
of \cite{jrh12} conclude that GTD fails to describe the thermodynamic properties of a phantom black hole. Indeed, classical black hole 
thermodynamics was established by assuming  that the laws of thermodynamics are also valid for black holes,  under the correct identification of the black hole thermodynamic variables, and consequently the properties of any particular black hole must be in agreement with the predictions of this approach. The formalism of GTD, on the contrary, is a geometric approach which uses the mathematical structure of classical thermodynamics in order to identify the equilibrium space as a Riemannian manifold whose geometric properties can be used alternatively to determine the thermodynamic properties of the corresponding physical system (for instance, phase transitions should correspond to curvature singularities). This implies that in case of divergent  results, one should assume as correct the predictions of classical black hole thermodynamics. 

The main goal of the present work is to show that the contradiction between classical thermodynamics and GTD, found in \cite{jrh12}, can be avoided by applying the fact that any thermodynamic system, even a phantom black hole, must be represented by a fundamental equation which is mathematically determined by a homogeneous function 
\cite{callen}.  
Indeed, we will show that for the fundamental function of the phantom black hole derived in \cite{jrh12} to be a homogeneous function, it is necessary to consider the cosmological constant as a thermodynamic variable. This implies that the corresponding equilibrium manifold must be 3-dimensional, instead of the 2-dimensional case 
investigated in \cite{jrh12}. Once the third thermodynamic dimension is taken into account the equivalence between classical black hole thermodynamics and GTD is recovered. Recently, using intuitive thermodynamic arguments, it has been suggested in \cite{dolan} that the cosmological constant should be considered as an intensive thermodynamic variable so that the enthalpy, instead of the mass, should correspond to the energy of a black hole. The analysis presented here can be interpreted as an additional more formal mathematical argument towards a possible proof of the thermodynamic nature of the cosmological constant.

This paper is organized as follows. In Sec. \ref{sec:pbh}, we present the explicit form and review the main properties of the anti-de-Sitter black hole with 
phantom charge. In Sec. \ref{sec:td}, we analyze the fundamental equation of the phantom black hole and derive its main thermodynamic properties, considering the cosmological constant as an additional thermodynamic variable. In Sec. \ref{sec:gtd}, we perform a geometrothermodynamic analysis of the 3-dimensional equilibrium manifold of the phantom black hole, and show that the structure of its thermodynamic curvature leads to results which are equivalent to the ones obtained from the analysis of the corresponding heat capacity. This proves the compatibility between classical black hole thermodynamics and GTD. Finally, in Sec. \ref{sec:con}, we discuss our results.


\section{An anti-de-Sitter black hole with phantom charge}
\label{sec:pbh}

Consider the Einstein-Hilbert action with cosmological constant $\Lambda$, minimally coupled to the electromagnetic field \cite{jrh12} (we will use geometric units throughout this work) 
\be
S= \int \sqrt{-g} \left (R+2\Lambda + 2\eta F_{\mu\nu}F^{\mu\nu}\right) d^4 x\ ,
\ee
where the constant $\eta$ indicates the nature of the electromagnetic field. For $\eta=1$, we obtain the classical Einstein-Maxwell theory, whereas 
for $\eta=-1$, the Maxwell field is phantom. The corresponding field equations are, in principle, equivalent to the Einstein-Maxwell system of differential equations. In particular, the constant $\eta$ enters only the Maxwell energy-momentum tensor at the level of the Einstein equations. 

In the case of spherically symmetric spacetimes
\be
ds^2 = f(r) dt^2 -\frac{1}{f(r)}dr^2 - r^2 (d\theta^2 + \sin^2\theta\, d\phi^2)\ ,
\label{pbhmet}
\ee
a particular solution is given by
\bea 
f(r) = 1 - \frac{2M}{r} -\frac{\Lambda}{3}r^2 + \eta\frac{q^2}{r^2}\ ,\nonumber \\
 F= \frac{1}{2} F_{\mu\nu}d x^\mu\wedge dx^\nu = \frac{q}{r^2} dt\wedge dr \ ,
\label{pbhsol}
\eea
where $q$ represents the electric charge of the source. This exact solution is regular in the entire spacetime, except at the origin $r=0$ where a curvature singularity exists. It is asymptotically flat for any finite values of the parameters $M$,  $q$ and $\Lambda=0$, and asymptotically anti-de-Sitter 
for $\Lambda\neq 0$. Moreover, 
it reduces to the Minkowski metric when all the parameters vanish. 
In this sense, this solution can be considered as physically meaningful. In the case of a phantom charge ($\eta=-1$), however, the energy contribution of the electromagnetic field to the action becomes negative and, therefore, it is interpreted as exotic matter. 

The solution (\ref{pbhmet})--(\ref{pbhsol}) corresponds to a black hole spacetime because of the presence of horizons. Indeed, the zeros of the function $f(r)$ 
determine the location of the horizons. In general, one can show that the equation $f(r)=0$ has two different roots which correspond to the radii of the horizons. 
In the case of a negative cosmological constant (anti-de-Sitter)  and a phantom charge ($\eta=-1$), only one of the zeros, say $r_+$, is positive. This implies that the condition 
\be
M= \frac{r_+}{2}\left(1 - \frac{\Lambda}{3} r_+^2 + \eta \frac{q^2}{r_+^2}\right) 
\label{mass}
\ee
must be satisfied on the horizon. For $\eta=1$, the case of the  AdS-Reissner-Nordstr\"om black hole is recovered.


\section{Fundamental equation and thermodynamics}
\label{sec:td}

Classical black hole thermodynamics \cite{dav77} is based upon the assumption of the validity of the laws of thermodynamics and the Hawking-Bekenstein entropy relation 
\be
S = \frac{1}{4} A\ ,
\label{hb}
\ee
where $A$ is the area of the outer horizon. In the case of spherically symmetric spacetimes $A=4\pi r_+^2$. Since the radius $r_+$ is obtained as a solution of the equation $f(r)=0$, it should depend only on the parameters entering the function $f(r)$. Then, Eq.(\ref{hb}) represents a relationship between the entropy $S$ and the parameters $M$, $\Lambda$ and $q$, i.e., it is a fundamental equation that relates all the parameters of the black hole. In concrete calculations, it is not always possible to solve explicitly the algebraic equation $f(r)=0$. Instead, it is always possible to use the identity (\ref{mass}) to find the fundamental equation which in this case reduces to 
\be
M= \frac{1}{2} S^{3/2} \left(\frac{1}{S} - \frac{\Lambda}{3} + \eta  \frac{q^2}{S^2}\right) \ ,
\label{feq}
\ee
where for the sake of simplicity we have rescaled the entropy as $S/\pi \rightarrow S$. 

According to classical black hole thermodynamics \cite{dav77a}, the fundamental equation should be given by a homogeneous function of the extensive variables. On the other hand, the mathematical definition of a homogeneous function of 
$n$ variables, say $\Phi(E^a),\ a=1,\ldots,n$, implies that $\Phi(\lambda E^a) = \lambda^\beta \Phi(E^a)$, where $\lambda$ and  $\beta$ are real constants, and $\beta$ 
indicates the degree of the homogeneous function. With $\Phi=M$ and $E^a=\{S,q\}$, it is easy to see that the function (\ref{feq}) is not homogeneous. However, the idea of homogeneous functions admits further generalization \cite{sta71}. Indeed, a function $\Phi(E^a)$ is called a generalized homogeneous function if it satisfies the condition
\be 
\Phi(\lambda^{\alpha_1} E^1,\ldots, \lambda^{\alpha_n}E^n) = \lambda^\beta \Phi(E^1,\ldots, E^n)\ ,
\ee
or, equivalently, $\Phi(\lambda^{\alpha_a}E^a) = \lambda^\beta \Phi(E^a)$, where $\alpha_1,\ldots, \alpha_n$ are arbitrary real constants. 
Notice that the degree of generalized homogeneous functions can always be set equal to one. Indeed, if a function satisfies the relationship  
$\Phi(\lambda^{\alpha_a}E^a) = \lambda^\beta \Phi(E^a)$,
it is possible to introduce a new constant $\bar \lambda = \lambda^\beta$ such that $\Phi(\bar \lambda ^{\alpha_a/\beta} E^a) = \bar \lambda \Phi(E^a)$, i.e., 
with a redefinition of the constant $\lambda$, the degree of $\Phi(E^a)$ is reduced to one. Alternatively, one can always introduce new coordinates 
$\bar E ^a =(E^a)^{\beta/\alpha_a}$ to obtain a first-degree generalized homogeneous function. 

Consider now the fundamental equation (\ref{feq}) with the rescaling 
$S\rightarrow \lambda^{\alpha_S}S $ and $q\rightarrow \lambda^{\alpha_q} q$. One can easily show that it is not a generalized homogeneous function, unless we consider 
$\Lambda$ as a thermodynamic variable which rescales as $\Lambda\rightarrow \lambda^{\alpha_\Lambda} \Lambda$. In this case, we obtain that if the conditions
\be
\alpha_\Lambda = - \alpha_S \ ,\qquad \alpha_q = \frac{1}{2}\alpha_S \ ,
\ee
are satisfied, 
the fundamental equation (\ref{feq}) is a generalized homogeneous function of degree $\beta= \alpha_S/2$. This proves that the cosmological constant must be considered as a thermodynamic variable in order for the phantom AdS black hole to be a thermodynamic system. Notice that in order for the mass to be an extensive variable (not necessarily linear), $\beta$ must be positive. Then, $\alpha_S>0$ and $ \alpha_\Lambda <0$, implying that $\Lambda$ is not an extensive variable. On the other hand, a fundamental equation like (\ref{feq}) should depend only on extensive variables. To ``convert" $\Lambda$ into an extensive variable, it is sufficient to consider its inverse. To conserve the physical meaning of the parameters entering the fundamental equation (\ref{feq}), we therefore introduce the radius of curvature $l$ as $l^2= - 3/\Lambda$, and obtain
\be
M=\frac{1}{2}S^{3/2} \left(\frac{1}{S} +\frac{1}{l^2} +\eta \frac{q^2}{S^2} \right)\ ,
\label{feq1}
\ee
where the radius of curvature $l$ rescales as $l\rightarrow \lambda^{\alpha_S/2} l$. This is now a fundamental equation which satisfies all the physical requirements. If we choose $\alpha_S=1$, it corresponds to a generalized homogeneous function of degree $1/2$, i.e.,
\be
M(\lambda S, \lambda^{1/2} l , \lambda^{1/2} q) = \lambda^{1/2} M(S,l,q)\ .
\ee
As mentioned before, the degree of any generalized harmonic function can be reduced to one, by choosing the variables appropriately. In the above case, it is sufficient to introduce the new entropy $s= S^{1/2}$ so that Eq.(\ref{feq1}) reduces to 
\be
m = \frac{1}{2} s^3  \left(\frac{1}{s^2} +\frac{1}{l^2} +\eta \frac{q^2}{s^4} \right)\ ,
\label{feq2}
\ee
which is, in fact, a first-degree homogeneous function. This is, however, a mathematical reduction which can change the physical properties of the system due to the transformation involving the power of the entropy (for this reason we use a different notation for the mass). We will see in the next section that using anyone of the fundamental equations (\ref{feq1}) or (\ref{feq2}), it is possible to avoid the contradictions found in \cite{jrh12}, when applying GTD to phantom black holes.

We will now explore the main thermodynamic properties of the phantom black hole under consideration in this work. First, let us note that the phantom mass function (\ref{feq1})  can become negative for certain values of the entropy and curvature radius. Let us assume that $S$ and $l$ are positive. Then, for $\eta=-1$, we obtain that there exists a critical charge 
$q_c=\sqrt{S(l^2+S)}/l$ at which the mass vanishes. For values $q>q_c$, the mass becomes negative. This exotic region appears as a result of the fact that the contribution of the electromagnetic energy is negative.  The first law of black hole 
thermodynamics in this case reads
\be 
dM = T dS + L dl + \eta A_0 d q\ ,
\label{flaw}
\ee
where $T$ is the temperature of the black hole, $L$ is the intensive variable dual to $l$, and $A_0$ is the electric potential which is in accordance with the Faraday 2-form (\ref{pbhsol}). The intensive variable $L=-(r_+/l)^3$ represents the ratio between the horizon and the curvature radii. Finally, the temperature
\be
T= \frac{1}{4S^{3/2} l^2}( S l^2 + 3 S^2 - \eta q^2 l^2) \ ,
\ee
is always positive for a  phantom charge ($\eta=-1)$, but can become negative otherwise. This means that a phantom black hole has no zero-temperature limit which, in this case, can be interpreted as an indication of the lack of phantom naked singularities.


\section{Geometrothermodynamic approach}
\label{sec:gtd}

The starting point of GTD is the thermodynamic phase space which is a $(2n+1)$-dimensional Riemannian contact manifold (${\cal T},\Theta, G)$, where ${\cal T}$ is a 
differential manifold, $\Theta$ is a contact form, i.e., $\Theta\wedge (d\Theta)^n \neq 0$,  and 
$G$ a Riemannian metric. If we introduce in ${\cal T}$ the coordinates $Z^A=\{\Phi, E^a, I^a\}$ with $a=1,\ldots n$
and $A=0,\ldots, 2n$, according to Darboux theorem, the contact form $\Theta$ can be expressed as $\Theta = d\Phi - \delta_{ab} I^a d E^b$. In order to reproduce 
the Legendre invariance (independence of the chosen thermodynamic potential) of classical thermodynamics, the metric $G$ must be 
invariant with respect to Legendre transformations \cite{arnold,quev07}. This is a problem  because, in general, Legendre transformations do not constitute 
a group and, therefore, it is not possible to apply the standard methods of differential geometry to derive the most general Legendre invariant metric \cite{gl14}.  
Nevertheless, it is possible to find particular solutions and impose physical requirements in order to derive quite general metrics which allows us to describe large classes of thermodynamic systems \cite{quev08,aqs08,qq11}. For instance, the metric
\be 
G = \Theta ^2 + (\delta_{ab} I^a E^b) (\eta_{cd} dE^c dE^d) \ ,
\label{GII}
\ee
has been used to describe black hole thermodynamics. Indeed, the idea is to induce a Legendre invariant metric $g = \varphi^*(G)$ for an $n-$dimensional 
submanifold ${\cal E}\subset {\cal T} $ by means of the pullback $\varphi^*$ which is associated with the smooth embedding map $\varphi: {\cal E} \rightarrow {\cal T}$, and satisfies the condition $\varphi^*(\Theta)=0$. If we choose the set $\{E^a\}$ as coordinates of ${\cal E}$, then the embedding reads 
$\varphi: \{E^a\} \mapsto \{\Phi(E^a), E^a, I^a(E^a)\}$ so that $\Phi(E^a)$ is the fundamental equation  and the induced metric becomes (up to a multiplicative constant)
\be
g= \varphi^*(G) = \Phi \, \eta^b_a \Phi_{,bc} \, d E^a dE^c\ ,
\label{gdown}
\ee
where $\eta_a^c ={\rm diag}(-1,1,\ldots,1)$. Notice that here we have used the Euler identity in the form $E^a \Phi_{,a} = \beta \Phi$ for homogeneous 
functions of degree $\beta$. 
However,  if $\Phi(E^a)$ is a generalized homogeneous function, the term $(\delta_{ab} I^a E^b)$ in Eq.(\ref{GII}) should be changed accordingly so that 
it induces the Euler identity in the form $\alpha_a E^a \Phi_{,a} = \alpha_\Phi \Phi$.  This issue has been discussed in detail in \cite{turco}.   

We now apply the above formalism to the case of phantom black holes. First, let us consider the first-degree homogeneous function (\ref{feq2}) which is mathematically equivalent to the fundamental equation (\ref{feq1}). In this case, the metric (\ref{gdown}) is 3-dimensional and can be expressed as
\be
g= m\left( -\frac{3s^4+\eta q^2l^2}{s^3l^2} ds^2 + \frac{\eta}{s} dq^2 + \frac{3s^3}{l^4} dl^2 \right)\ , 
\label{metfeq2}
\ee  
which leads to the following scalar curvature 
\be
R = \frac{4 N(s,q,l)}{3 (s^4 + s^2 l^2 +\eta q^2 l^2)^3(3s^4+\eta q^2 l^2 )^2}\ ,
\ee
\bea
N(s,q,l)= & 9\,{s}^{8} \left( 7\,{s}^{2}+{l}^{2} \right)  \left( {s}^{2}+{l}^{2}
 \right) -{s}^{2}{q}^{4}{l}^{4} \left( 4\,{l}^{2}+31\,{s}^{2} \right) \nonumber\\
&-3\,\eta\,{s}^{4}{q}^{2}{l}^{2} \left( 23\,{s}^{4}-4\,{s}^{2}{l}^{2}+3
\,{l}^{4} \right) +5\,\eta\,{q}^{6}{l}^{6} \ ,\\
\eea
where we have used the relation $\eta^2 =1$ to simplify the expressions. We see that there are two curvature singularities. The first one occurs when the condition
$(s^4 + s^2 l^2 +\eta q^2 l^2)=0$ is satisfied, which has real solutions in the phantom case. However, it is easy to see that this condition implies also that $m=0$,
which can be interpreted as unphysical, because in this case the first law of thermodynamics breaks down and the thermodynamic metric $g$ is not defined at all. 
The second singularity implies that $3s^4 =-\eta q^2 l^2$, which is valid only in the case of phantom black holes. According to the conjectures of GTD, this means that 
this phantom black hole undergoes a second-order phase transition, once the phantom charge is given by $q=\pm\sqrt{3} s^2/l$. To see if the theory of black hole thermodynamics 
predicts a similar behavior, we calculate the heat capacity \cite{dav77}
\be
C_{q,l} = T\left(\frac{\partial S}{\partial T}\right) = 
\frac{s(3s^4+s^2 l^2 -\eta q^2 l^2)}{2(3s^4 + \eta q^2 l^2)} \ ,
\ee
which indicates a second-order phase transition for $(3s^4 + \eta q^2 l^2)=0$. This coincides with the curvature singularity of the metric (\ref{metfeq2}) and proves
the compatibility between the results of GTD and black hole thermodynamics. 
 
Consider now the fundamental equation (\ref{feq1}) which, according to the analysis presented above, is a generalized homogeneous function of degree $1/2$ that does not involve a redefinition of the thermodynamic variables, affecting the physical properties of the thermodynamic system. 
The thermodynamic metric (\ref{gdown}) is again 3-dimensional and reduces to
\be
g= M\left( -\frac{3S^2 - Sl^2 + 3 \eta q^2 l^2}{8S^{5/2} l^2} dS^2 + \frac{\eta}{S^{1/2}} d q^2 + 3 \frac{S^{3/2}}{l^4} d l^2\right) \ .
\ee
Notice that in this case the Euler identity, which generates the conformal factor of $g$, reads $2 M= 2 S T + l L + \eta q A_0$, using the notations introduced in 
Eq.(\ref{flaw}). A straightforward calculation leads to the
following curvature scalar 
\be
R = \frac{8S l ^4 N(S,q,l)}{3(S^2 + S l^2 + \eta q^2 l^2)^3 (3S^2-Sl^2 + 3 \eta q^2 l^2)^2} \,
\label{curv2}
\ee
where
\bea 
N(S,q,l)= & {S}^{3} \left( 4\,{l}^{6}+36\,{S}^{3}+7\,S{l}^{4}+27\,{S}^{2}{l}^{2}
 \right) -9\,{l}^{4}{q}^{4}S \left( {l}^{2}+12\,S \right) \nonumber\\
& + \eta \left[18\,{q}^{6}{l}^{6} -{q}^{2}{l}^{2}{S}^{2} \left( 59\,{l}^{4}+18\,S{l}^{2}+90\,{S}^{2}
 \right) \right] \ .
\eea
The first curvature singularity located at $(S^2 + S l^2 + \eta q^2 l^2)=0$ implies that the mass of the black hole vanishes; as mentioned above, we interpret this type of singularities as unphysical. The second singularity is determined by the zeros of  $(3S^2-Sl^2 + 3 \eta q^2 l^2)$, and indicates that there exists a second-order phase transition in phantom black holes that satisfy the condition $l=\pm 3S /\sqrt{3S + 9 q^2}$.  This is the geometric interpretation of a phase transition according to GTD. Now, we will see that this result is compatible with the interpretation of a phase transition in black hole thermodynamics. Indeed, the  heat capacity that corresponds to the fundamental equation (\ref{feq1}) is given by
\be
C_{q,l} = \left(\frac{\partial M}{\partial S}\right) \left( \frac{\partial^2 M}{\partial S^2} \right)^{-1} = \frac{ 2 S ( Sl^2+ 3S^2 - \eta q^2 l^2)}
{3S^2 - Sl^2 + 3\eta q^2 l^2} \ .
\ee  
We see that the phase transitions occur for $(3S^2 - Sl^2 + 3\eta q^2 l^2)=0$, a condition that determines exactly the same points of the equilibrium space where the thermodynamic curvature (\ref{curv2}) diverges.

\section{Final remarks}
\label{sec:con}

In this work, we investigated the thermodynamics and geometrothermodynamics of a spherically symmetric AdS charged phantom black hole. The electric charge has a phantom nature because the electromagnetic energy density that enters the corresponding action is negative. As a consequence, the field equations allow the existence of exact solutions with exotic properties. 
First, we investigated the fundamental equation
that relates the total mass, the entropy and the phantom charge.  
We proved that for the fundamental equation to be a homogeneous function, it is necessary to consider the cosmological constant as a thermodynamic variable. It turns out 
that the cosmological constant has the properties of an intensive variable and, therefore, in order for the fundamental equation to depend on extensive variables only, it is necessary to use the curvature radius instead. As a result, we obtain a fundamental equation 
whose mathematical properties resemble those of classical thermodynamic systems. Considering the cosmological constant as a thermodynamic variable implies that the equilibrium space must be extended by one dimension. A similar result was obtained recently \cite{dolan}, by assuming that the energy of a black hole is not represented by its total mass, but by the corresponding enthalpy, indicating that the cosmological constant is an intensive thermodynamic variable similar to the pressure. Our results corroborate from a more formal mathematical point of view the intuitive analysis performed in \cite{dolan}.

We investigate the properties of the extended 3-dimensional equilibrium space in the framework of GTD. To this end, we perform two different analysis. First, we show that it is possible to redefine the entropy of phantom black holes in such a way that the fundamental equation turns out to be a first-degree homogeneous function. In this case, the resulting physical properties of the system can be modified due to the redefinition of a thermodynamic variable. Nevertheless, we perform a comparison between the geometric properties of the equilibrium space and the predictions of the theory of black hole thermodynamics. We show that there exist curvature singularities at those places in the equilibrium space where second-order phase transitions occur. We interpret this result as a mathematical proof that GTD can 
correctly describe the properties of particular phantom black holes. 

A more physical approach consists in analyzing the fundamental equation as a generalized homogeneous function of degree 1/2, without redefining thermodynamic variables.
In this case, we also investigate the main thermodynamic properties of the phantom black hole system, and show that the analysis of the corresponding extended equilibrium space in the framework of GTD is compatible with the predictions of classical black hole thermodynamics. We conclude that GTD  correctly describes the 
thermodynamic properties of this particular class of AdS black holes with phantom charge.

\section*{Acknowledgements}

This work was carried out within the scope of the project CIAS 1790
supported by the Vicerrector\'\i a de Investigaciones de la Universidad
Militar Nueva Granada - Vigencia 2015. 
This work was partially supported
by DGAPA-UNAM, Grant No. 113514, and CONACyT, Grant No. 166391.

\end{document}